\documentclass[twocolumn,aps]{revtex4-1}

\usepackage{graphicx} 

\begin{document}

\title{Optimal complexity correction of correlated errors in the surface code}

\author{Austin G. Fowler$^{1,2}$}

\affiliation{$^1$Department of Physics, University of California, Santa Barbara, California 93106,
USA \\
$^2$Centre for Quantum Computation and Communication
Technology, School of Physics, The University of Melbourne, Victoria
3010, Australia}

\date{\today}

\begin{abstract}
The surface code is designed to suppress errors in quantum computing hardware and currently offers the most believable pathway to large-scale quantum computation. The surface code requires a 2-D array of nearest-neighbor coupled qubits that are capable of implementing a universal set of gates with error rates below approximately 1\%, requirements compatible with experimental reality. Consequently, a number of authors are attempting to squeeze additional performance out of the surface code. We describe an optimal complexity error suppression algorithm, parallelizable to $O(1)$ given constant computing resources per unit area, and provide evidence that this algorithm exploits correlations in the error models of each gate in an asymptotically optimal manner.
\end{abstract}

\maketitle

\begin{figure}
\begin{center}
\resizebox{60mm}{!}{\includegraphics{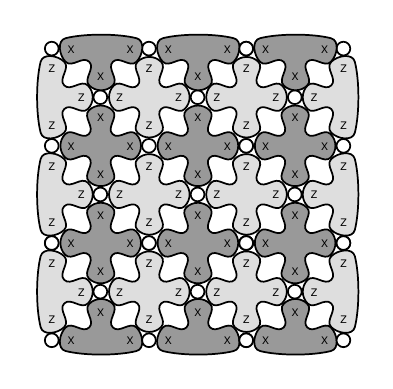}}
\end{center}
\caption{Distance 4 surface code. White circles represent data qubits. Each shaded bubble represents an $X$ or $Z$ stabilizer \cite{Gott97}, and the error-free surface code state can be thought of as the simultaneous +1 eigenstate of all stabilizers. When errors occur, the surface code state becomes the -1 eigenstate of some stabilizers. At least $\lceil d/2 \rceil$ errors must occur for stabilizer measurements, which report whether the underlying state is the +1 or -1 eigenstate of that stabilizer, to be ambiguous and have the potential to lead to a logical error after correction.}\label{sc}
\end{figure}

\begin{figure}
\begin{center}
\resizebox{80mm}{!}{\includegraphics{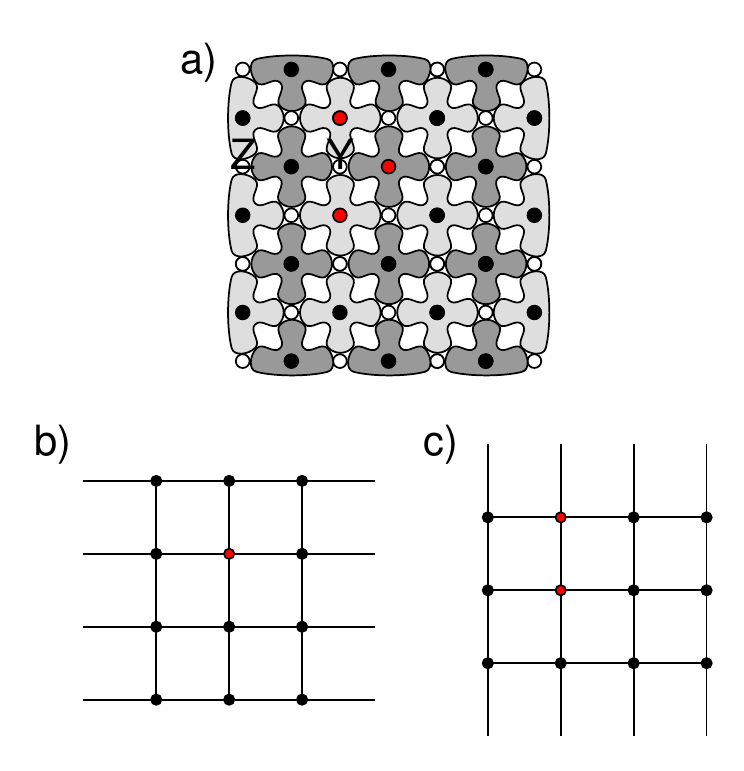}}
\end{center}
\caption{(Color online) a) Distance 4 surface code with two data qubit errors. Stabilizers containing light dots will report -1 eigenstates when measured (indicating a detection event). b) The $X$ stabilizer measurements can, with reasonable effectiveness, be used in isolation to predict the location of $Z$ errors. The minimum weight perfect matching algorithm takes the indicated structure as input, and outputs a pairing of the detection events to one another or a boundary making use of the minimum total number of edges. In this ambiguous case however, the two edges to the left or right of the single detection event will be returned with equal probability, resulting in corresponding $Z$ corrections and success or failure, respectively. c) The $Z$ detection events will be matched with one another, leading to successful correction of the $X$ component of the $Y$ error.}\label{prob}
\end{figure}

The surface code \cite{Brav98,Denn02,Raus07,Raus07d,Fowl08,Fowl12f} is structurally simple, involving a 2-D array of qubits and the measurement of local operators (stabilizers \cite{Gott97}) as shown in Fig.~\ref{sc}. If we assume that stabilizers can be measured without error, a given pattern of random $X$, $Y$, $Z$ errors on some subset of the data qubits can be written as two 2-D graph problems (Fig.~\ref{prob}). Given only nearest neighbor gates and no perfect operations, one can construct a similar pair of 3-D graph problems.

In extensive prior work, we have processed the two problems indicated in Fig.~\ref{prob} independently \cite{Wang09,Fowl10,Fowl11b,Fowl12c,Fowl12g,Fowl13d}. Other authors have already considered exploiting knowledge of correlations between $X$ and $Z$ errors to lower the probability of unsuccessful correction. These authors have assumed errors on single data qubits only, perfect stabilizer measurements, and in some cases have also considered a probability of classical bit-flip on the reported stabilizer measurement result \cite{Ducl09,Ducl10,Ducl13,Woot12,Woot13,Bomb12}. To date, however, no author has considered exploiting error correlations when given only a 2-D array of qubits with nearest neighbor interactions and no perfect gates. In this work, we present simulations of this realistic case, prove that our runtime is complexity optimal, and provide evidence that our performance is asymptotically optimal given a depolarizing error rate $p\lesssim 2\times 10^{-4}$.

The discussion is organized as follows. In Section~\ref{perf1}, we present simulations of the perfect stabilizer measurement case both without exploiting correlations and with the simplest imaginable approach to dealing with correlations. We furthermore show that this simple approach is close to optimal. In Section~\ref{full}, we extend our approach to the fully fault-tolerant case, present simulation data, and prove that the runtime is complexity optimal. Section~\ref{conc} concludes.

\section{Perfect stabilizer measurements}
\label{perf1}

\begin{figure}
\begin{center}
\resizebox{85mm}{!}{\includegraphics[viewport=60 60 545 430, clip=true]{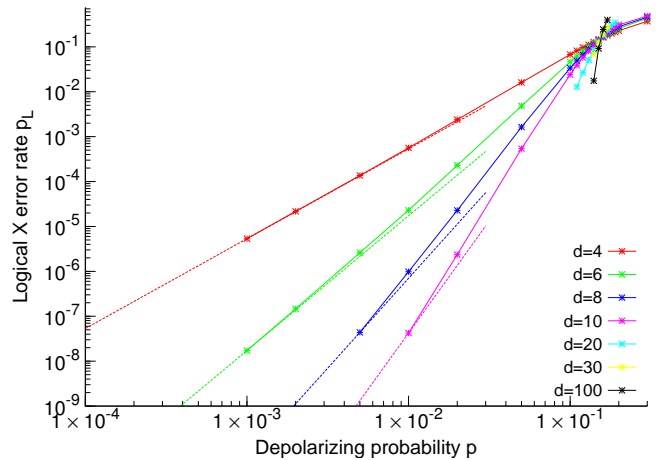}}
\end{center}
\caption{(Color online) Basic perfect stabilizer measurement correction. Probability of logical $X$ error $p_L$ as a function of the depolarizing error probability $p$ for a range of distances $d=4, \ldots, 100$ when performing basic minimum weight perfect matching only. Referring to the left of the figure, the distance increases top to bottom. Quadratic, cubic, quartic, and quintic lines (dashed) have been drawn through the lowest distance 4, 6, 8, 10 data points obtained, respectively.}\label{logx_u}
\end{figure}

We first need a baseline of the performance of independent minimum weight perfect matching of the two problems shown in Fig.~\ref{prob}. This baseline graph of the probability of correction failure versus the probability of physical error can be found in Fig.~\ref{logx_u}. We show only the probability of correction failure due to $X$ errors as the $Z$ case is identical by symmetry. Our goal in this Section is to devise a better algorithm resulting in a lower probability of correction failure.

The edges in the graphs of Figs.~\ref{prob}b--c are associated with equal weights $w=-\ln p$. After obtaining minimum weight perfect matchings of these two graphs, we focus on matched pairs of detection events, or a single detection event attached to a boundary, connected by only one edge. At low error rates, these are the dominant class of matchings. Such a matched edge indicates the high probability presence of either an $X$ or $Z$ error, depending on which problem is being solved. Referring to Fig.~\ref{prob}c, we shall for discussion purposes focus on the $X$ error located with high confidence by matching this problem. The high confidence of an $X$ error in this position allows us to infer that there is a high probability of a $Z$ error as well due to the depolarizing noise model. We therefore reweight the edge to the left to the detection event in Fig.~\ref{prob}c to $w=-\ln 0.5$. If there had been any single-edge matchings in Fig.~\ref{prob}b we would similarly reweight the corresponding perpendicular edge in Fig.~\ref{prob}c.

\begin{figure}
\begin{center}
\resizebox{85mm}{!}{\includegraphics[viewport=60 60 545 430, clip=true]{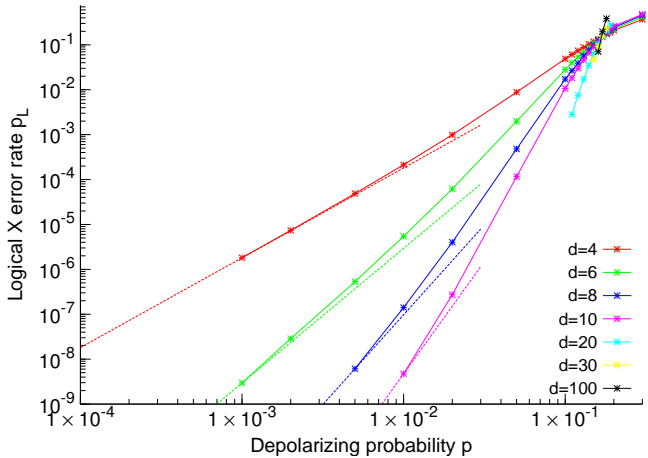}}
\end{center}
\caption{(Color online) Perfect stabilizer measurement correlated error correction. Probability of logical $X$ error $p_L$ as a function of the depolarizing error probability $p$ for a range of distances $d=3, \ldots, 100$ when performing two rounds of minimum weight perfect matching with edge reweighting based on the first round. Referring to the left of the figure, the distance increases top to bottom. Quadratic, cubic, quartic, and quintic lines (dashed) have been drawn through the lowest distance 4, 6, 8, 10 data points obtained, respectively.}\label{logx_c}
\end{figure}

After reweighting, minimum weight perfect matching is performed again, and in this case the reweighted edge allows us to correctly infer that the chain of $Z$ errors running from the central detection event in Fig.~\ref{prob}b runs to the left boundary. The error suppression performance of this simple augmentation of minimum weight perfect matching is shown in Fig.~\ref{logx_c}.

Consider any top to bottom non-intersecting length $n$ pathway. If $\lceil n/2 \rceil$ errors with $X$ components occur along this pathway, there is the potential for a logical error. Given
\begin{equation}
\label{limit}
\begin{array}{c}{\rm lim} \\ n\rightarrow\infty \end{array} \left(\begin{array}{c}n \\ \lceil n/2 \rceil \end{array}\right)/\left(\begin{array}{c}n \\ \lceil n/2 \rceil +1 \end{array}\right) = 1,
\end{equation}
we have that for large $d$, which sets a minimum value of $n$, and \emph{fixed} small $p$, the contribution of patterns of more than the minimum number of errors along a given pathway to the total logical error rate is negligible. Note that this does not preclude $n>d$, it just means we can restrict our attention to $\lceil n/2 \rceil$ errors along any given path of length $n$. The value of $p$ at which this approximation is valid for the perfect stabilizer measurement case is approximately $p_c\sim 10^{-2}$, corresponding to an approximately 1\% contribution to the logical error rate by higher numbers of errors along a given pathway.

At any fixed $d$, for sufficiently low $p$, lowest order $n=d$ effects dominate and the logical error rate scales as $O(p^{d/2})$. Referring to Fig~\ref{sc}, minimum length logical $X$ operators must run straight from top to bottom starting on one of the data qubits in the top row. There are $d$ such minimum length logical $X$ operators in a distance $d$ square surface code. Basic matching of detection events associated with $X$ errors will fail 50\% of the time if $d/2$ locations along a minimum length logical operator each contain an error with an $X$ component. The low distance, low physical error first order probability of logical error is therefore
\begin{equation}
\label{exact}
p_L=\frac{d}{2}\left(\begin{array}{c}d \\ d/2 \end{array}\right)(2p/3)^{d/2}.
\end{equation}
For $d=4$ and $p=10^{-3}$, Eq.~\ref{exact} gives $p_L=5.3\times 10^{-6}$ and $p_L=1.7\times 10^{-8}$ for $d=6$. These values match the simulated data to two significant figures, providing solid verification of both our theoretical understanding and the correctness of our simulations.

\begin{figure}
\begin{center}
\resizebox{80mm}{!}{\includegraphics{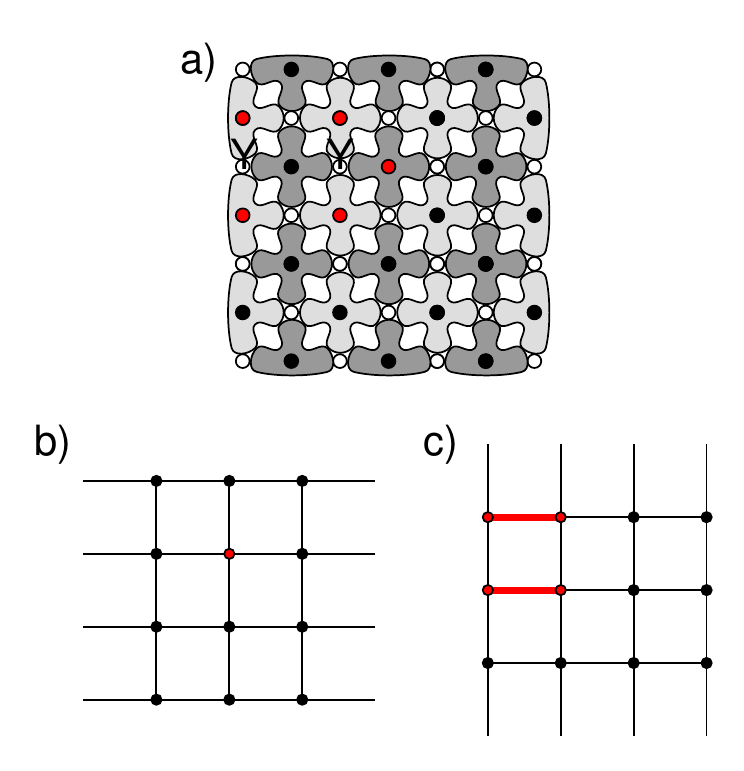}}
\end{center}
\caption{(Color online) a) Distance 4 surface code with two data qubit errors. Stabilizers containing light dots will report -1 eigenstates when measured (indicating a detection event). b) The minimum weight perfect matching algorithm will return the two edges to the left or right of the single detection event with equal probability, resulting in success or failure, respectively. c) The $Z$ detection events can be matched as shown, leading to no helpful reweighting of b).}\label{prob2}
\end{figure}

Ideally, taking correlations into account should ensure successful correction if there are any $Y$ errors, for $n$ even, and at least two $Y$ errors, for $n$ odd. For low $d$, low $p$ and $d$ even, the ideal logical error rate should therefore be
\begin{equation}
\label{perfect}
p_L=\frac{d}{2}\left(\begin{array}{c}d \\ d/2 \end{array}\right)(p/3)^{d/2}.
\end{equation}
No method of correction can perform better than Eq.~\ref{perfect} in the stated parameter range. An algorithm capable of achieving this error scaling assuming perfect stabilizer measurements was first presented in \cite{Woot13}. The simple method of correlated correction described can, however, fail more often than this due to the matching of even length sequences of $Y$ errors not having a unique solution (see Fig.~\ref{prob2} for an example where this leads to no helpful reweighting). It might seem that there is an obvious solution to this, namely taking into account when there are multiple minimum weight perfect matchings, however we shall see that this is not necessary.

If the potential logical error contains at least one odd length chain of $Y$ errors, helpful reweighting will occur and the logical error will be avoided. For code distances likely to be used in a practical quantum computer, say $d=20$ and hence $n\geq 20$, the probability of a potential logical error containing no odd length chains of $Y$ errors is quite small. To be precise, there are 184,756 ways of choosing 10 error locations out of 20 positions, and for each pattern there are 1024 ways of assigning $X$ and $Y$ errors, yet of all these approximately $1.9\times 10^8$ error strings, only $2.3\times 10^6$, or 1.2\%, do not contain at least one odd length chain of $Y$ errors. Note furthermore that even if there are no odd length chains of $Y$ errors, the probability that all of the detection events associated with the $Z$ errors are matched in a manner not crossing the potential logical error pathway is low. Similar behavior is observed for $n$ odd, and the probability of unfavorable reweighting decreases with increasing $n$.

\begin{figure}
\begin{center}
\resizebox{40mm}{!}{\includegraphics{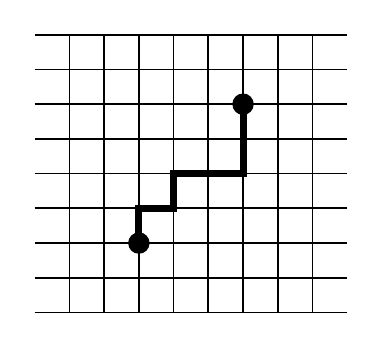}}
\end{center}
\caption{Distance 9 code of the form shown in Fig.~\ref{sc}, line intersections represent the location of $X$ stabilizers. Example of a 7 $Z$ error chain that a simple minimum weight matching will fail to correct, leading to a logical error. Note that the detection events can only be matched to the left and right boundaries.}\label{classes}
\end{figure}

We also need to consider the fact that a minimum weight matching is not always the best thing to do. Consider Fig.~\ref{classes}. This shows an error chain of length 7 and two resultant detection events. Based on this information, a minimum weight matching will lead to each detection event being matched to its nearest boundary. However, there are 35 different length 7 pathways connecting the two detection events, for a leading order total probability of $35p^7$. By contrast, there is only one total length 6 matching to the boundaries, and 12 length 7 boundary matchings, for a total probability of $p^6+12p^7$, neglecting higher order terms. For $p>1/23$, pairing the detection events is therefore a better choice than connecting them to their nearest boundaries. Note that the example given in Fig.~\ref{classes} is large and rare, and that $p=1/23$ is a high error rate. Smaller examples require even higher values of $p$ to trigger. At fixed small $p$, it takes very widely separated detection events with many possible paths between them to make a matching other than a minimum weight matching the preferred option. Furthermore, if two of the errors along the chain contain a component of the opposite type, after reweighting a minimum weight perfect matching will correctly choose to connect the two detection events.

The contribution of $n>d$ paths to the logical error rate will mean that at any fixed value of $p$ the logical error rate will tend to be above that shown in Eq.~\ref{perfect}. Nevertheless, given the discussion above, we conjecture that the above described algorithm will achieve logical error rates a vanishingly small constant above that achievable by an optimal algorithm at fixed $p\lesssim 10^{-2}$ in the limit of large $d$, with practical indistinguishability occurring for $d\gtrsim 20$.

\section{Full fault-tolerance}
\label{full}

\begin{figure}
\resizebox{60mm}{!}{\includegraphics{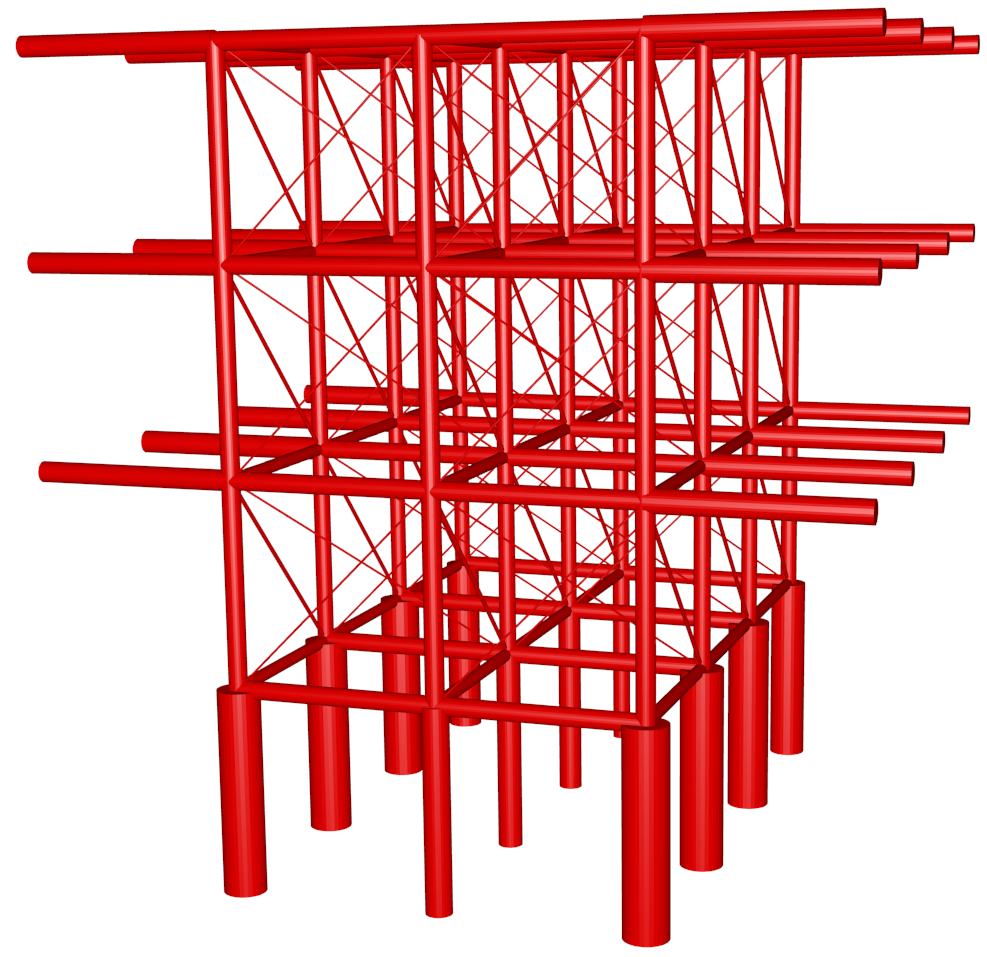}}
\caption{(Color online) Autotune generated $Z$ error detection ($X$ stabilizer measurement) 3-D graph problem associated with a distance 4 surface code with depolarizing noise. Each vertical cylinder can be thought of as being associated with a round of stabilizer measurement. If two temporally sequential stabilizer measurements disagree in value, a detection event is associated with the junction where the associated cylinders meet. The diameter of each cylinder is proportional to the total probability of all possible single gate errors (including two-qubit correlated errors) that will lead to detection events at the endpoint of the cylinder. Given a pattern of detection events, each cylinder is associated with a weight $w=-\ln p_{\rm cylinder}$ and minimum weight perfect matching used to find an appropriate set of cylinders connecting detection events in pairs or to boundaries with minimum total weight.}
\label{Autotune_sc}
\end{figure}

\begin{figure}
\begin{center}
\resizebox{85mm}{!}{\includegraphics[viewport=60 60 545 430, clip=true]{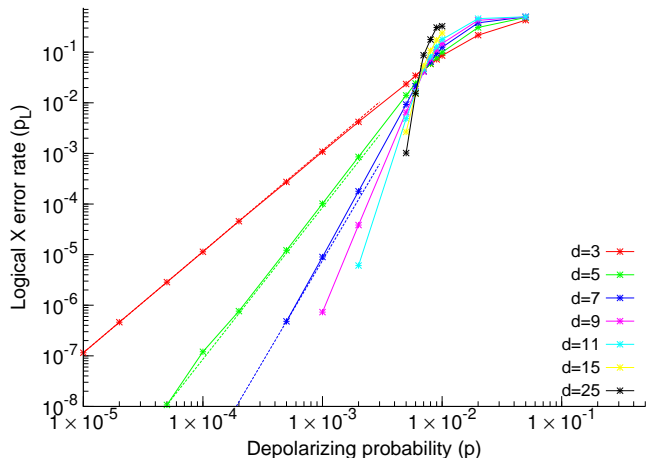}}
\end{center}
\caption{(Color online) Basic fault-tolerant correction. Probability of logical $X$ error per round of fault-tolerant error detection $p_L$ as a function of the depolarizing error probability $p$ for a range of distances $d=3, \ldots, 25$ when performing basic minimum weight perfect matching only. Referring to the left of the figure, the distance increases top to bottom. Quadratic, cubic, and quartic lines (dashed) have been drawn through the lowest distance 3, 5, 7 data points obtained, respectively.}\label{logx_ft_u}
\end{figure}

When simulating fully fault-tolerant stabilizer measurement, namely making use of only a 2-D array of qubits with nearest neighbor interactions and no perfect gates, we use Autotune \cite{Fowl12d} to determine where and when every possible error on every gate will be detected and the total probability of every possible pair of detection events arising from single errors. This leads to two 3-D graph problems, an example of which is shown in Fig.~\ref{Autotune_sc}. The baseline performance of independent minimum weight perfect matching of these two problems is shown in Fig.~\ref{logx_ft_u}.

\begin{figure}
\begin{center}
\resizebox{85mm}{!}{\includegraphics[viewport=60 60 545 430, clip=true]{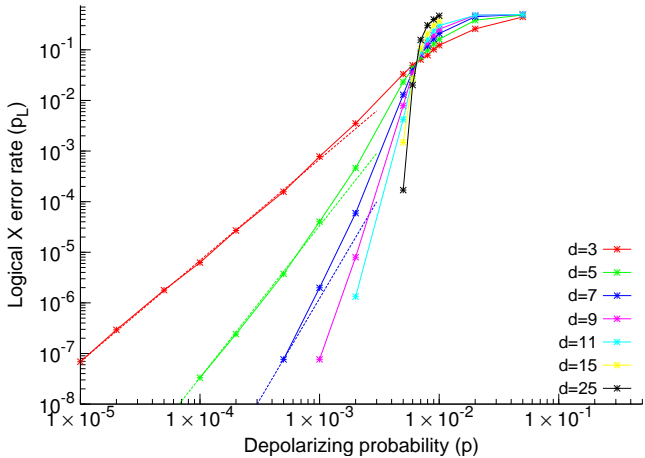}}
\end{center}
\caption{(Color online) Correlated fault-tolerant correction. Probability of logical $X$ error per round of fault-tolerant error detection $p_L$ as a function of the depolarizing error probability $p$ for a range of distances $d=3, \ldots, 25$ when exploiting knowledge of correlations between $X$ and $Z$ errors. Referring to the left of the figure, the distance increases top to bottom. Quadratic, cubic, and quartic lines (dashed) have been drawn through the lowest distance 3, 5, 7 data points obtained, respectively.}\label{logx_ft_c}
\end{figure}

To move beyond independent matching, consider, in analogy to the perfect stabilizer case, a pair of $Z$ error detection events we have matched using a single cylinder. This cylinder contains a list of all the specific errors occurring on specific gates that can result in this particular pair of detection events. If we choose to believe with certainty that one of these errors occurred, we can uniformly scale up the probability of each of these errors so that they total to 1. Some of these errors will contain $X$ components, allowing us to reweight the appropriate cylinders in the $X$ error 3-D graph problem, and then perform a second round of matching. The performance of this approach is shown in Fig.~\ref{logx_ft_c}.

The surface code provably exponentially suppresses error with increasing code distance $d$ at a sufficiently low fixed error rate below threshold \cite{Fowl12e}. We can therefore look at the ratio of distance 3 and 5 logical error rates at low $p$ and extrapolate with reasonable accuracy to higher distances. Attempts have been made recently to more accurately predict high distance logical error rates \cite{Brav13}. At $p=10^{-4}$ the ratio of distance 3 and 5 logical error rates is approximately 95 with independent matching, and 188 with correlated matching. Similarly, at $p=2\times 10^{-4}$ the ratio of distance 3 and 5 logical error rates is approximately 60 with independent matching, and 112 with correlated matching. Both of these indicate an approximate factor of 2 improvement from taking error correlations into account. We conjecture that, in analogy to the perfect stabilizer measurement case, the performance of the fully fault-tolerant algorithm described above will approach that of an optimal algorithm in the limit of large $d$ at fixed $p\lesssim 2\times 10^{-4}$.

In \cite{Fowl13e} we formally proved that minimum weight perfect matching can be performed in $O(1)$ time using a uniform 2-D array of processes communicating with their nearest neighbors and other physically reasonable assumptions. Unlike the 2-D graph problem associated with perfect stabilizer measurement, the 3-D graph problem associated with fully fault-tolerant stabilizer measurement must be solved continuously as new data is generated, not resolved from scratch if we reweight edges. Fortunately, it is provably exponentially unlikely that data far in the past will be considered when matching a given vertex \cite{Fowl13e}, and hence when reweighting we only need to unmatch detection events local to reweighted edges, introducing a constant average additional processing cost per detection event. In practice, the reweighting cost is negligible a low error rates, and the overall cost of taking correlations into account is approximately a factor of 2 slower processing due to the need to perform approximately twice as much matching, leaving the complexity at $O(1)$ parallel, which is optimal for any algorithm.

\section{Conclusion}
\label{conc}

We have described and benchmarked the error suppression performance of an optimal complexity algorithm taking a distance $d$ surface code and achieving logical error rates an approximate factor of $2^{d/2}$ lower than basic minimum weight perfect matching when assuming a sufficiently low error balanced depolarizing channel. Both perfect stabilizer measurement and fully fault-tolerant simulations have been presented, the latter assuming no perfect operations and only a 2-D array of qubits with nearest neighbor interactions. Arguments have been presented suggesting that the fully fault-tolerant algorithm's error suppression performance will be asymptotically optimal at any depolarizing error rate $p$ below a fixed and likely long-term experimentally achievable value $p_c\sim 2\times 10^{-4}$ and sufficiently large $d$. The performance boost decreases smoothly as the hardware error rate approaches the threshold error rate, which is itself also slightly improved by our algorithm. Our algorithm is built on top of our Autotune software \cite{Fowl12d}, meaning it can handle both arbitrary underlying hardware and arbitrary topological quantum error correction schemes that can be decoded using standard minimum weight perfect matching.

In future work, we will focus on the continued improvement of the runtime of our algorithm, and implementation of this algorithm in hardware for integration with experiments.

\section{Acknowledgements}
\label{ack}

We thank Sergey Bravyi, Ruben Andrist, and James Wootton for helpful comments. This research was funded by the US Office of the Director of National Intelligence (ODNI), Intelligence Advanced Research Projects Activity (IARPA), through the US Army Research Office grant No. W911NF-10-1-0334. Supported in part by the Australian Research Council Centre of Excellence for Quantum Computation and Communication Technology (CE110001027) and the U.S. Army Research Office (W911NF-13-1-0024). All statements of fact, opinion or conclusions contained herein are those of the authors and should not be construed as representing the official views or policies of IARPA, the ODNI, or the US Government.

\bibliography{../References}

\end{document}